\documentclass[reqno,12pt]{article} 
\usepackage{amsmath,amssymb,amsfonts}
\newcommand{\ov}{\overline}

\newcommand{\cH}{{\cal H}}

\newcommand{\nn}{\nonumber}

\newcommand{\bR}{{\bf R}}
\newcommand{\bT}{{\bf T}}

\newcommand{\wh}{\widehat}

\newcommand{\vp}{\varphi}

\begin{document}

\title{Dependence of the superfluidity criterion 
on the resonance between one-particle (Bogoliubov)
and two-particle series}
\author{V. P. Maslov}
\date{}

\maketitle

\begin{abstract}
We study how the superfluidity depends on the resonance
between one- and two-particle series.
The frequency of the spectrum of two-particle solutions 
in an interval is calculated. 
\end{abstract}

We consider a system of $N$ identical bosons of mass $m$
that lie on a three-dimensional torus $\bT$
with the side lengths $L_1$, $L_2$, and $L_2$. 
We assume that the bosons mutually interact and that the
interaction potential has the form 
\begin{equation}
V\left(N^{1/3}(x-y)\right),
\label{1aa}
\end{equation}
where $V(\xi)$ is an even smooth function 
rapidly decreasing at infinity and $x,y$ are the boson
coordinates on $\bT$. 
The interaction potential~(\ref{1aa})
depends on $N$ such that the interaction radius 
decreases as the particle number $N$ increases, 
and the average number of particles with which a single particle
interacts remains constant in this case.

In the ultrasecond quantization over pairs,
an ultrasecond-quantized operator~$\ov{\wh{H}}$,
whose explicit form was previously presented, 
corresponds to the boson system in question.
As stated, this ultrasecond-quantized operator 
satisfies identity 
\begin{equation}
\ov{\wh{H}}=\ov{\wh{E}}\wh{A},
\label{2aa}
\end{equation} 
where $\ov{\wh{E}}$ is the ultrasecond-quantized unit operator
and $\wh{A}$ is an operator nonuniquely chosen 
in the ultrasecond quantization space.
Further, we assume that the operator $\wh{A}$ has the form 
\begin{eqnarray}
&&\wh{A}=\iint dxdy\wh{B}^+(x,y)\left(-\frac{\hbar^2}{2m}(\Delta_x+\Delta_y)
+V\left(N^{1/3}(x-y)\right)\right)\wh{B}^-(x,y)+\nn\\
&&+2\int dxdydx'dy' V\left(N^{1/3}(x-y)\right)\wh{B}^+(x,y)\wh{B}^
+(x',y')\wh{B}^-(x,x')\wh{B}^-(y,y'),
\label{3aa}
\end{eqnarray}
where $\wh{B}^+(x,y)$ and $\wh{B}^-(x,y)$ are the corresponding 
bosonic creation and annihilation operators  
for a particle pair in the Fock ultrasecond quantization space.
By identity~(\ref{2aa}), 
an asymptotic expression for operator~(\ref{3aa})
can be found for establishing an asymptotic representation 
for the spectrum of the boson system under consideration 
in the limit as $N\to\infty$.

Because the function~(\ref{1aa}) times $N$
in a sense converges to the Dirac function 
in the limit as $N\to\infty$, the operator~(\ref{3aa}) 
involves the small parameter $1/N$
in the second term for this limiting case. 
This means that to find the asymptotic expressions 
for eigenfunctions and eigenvalues of the operator $\wh{A}$,
we can use the semiclassical methods developed in \cite{VKB}. 
The asymptotic expressions for eigenfunctions and eigenvalues 
are determined by the symbol of the operator~(\ref{3aa}), 
and this symbol is called the true symbol 
for the ultrasecond-quantized problem. 
The true symbol corresponding to the operator~(\ref{3aa})
is the following functional defined for a pair of functions
$\Phi^+(x,y)$ and $\Phi(x,y)$:
\begin{eqnarray}
&&\cH\left[\Phi^+(\cdot),\Phi(\cdot)\right]=\iint dxdy\Phi^+(x,y)
\left(-\frac{\hbar^2}{2m}(\Delta_x+\Delta_y)\right)\Phi(x,y)+\nn\\
&&+2\int dxdydx'dy' \left(NV\left(N^{1/3}(x-y)\right)\right)\Phi^+(x,y)
\Phi^+(x',y')\Phi(x,x')\Phi(y,y').
\label{4aa}
\end{eqnarray}
The conservation of the number of particles in the system 
for the functions $\Phi^+(x,y)$ and $\Phi(x,y)$
implies the condition
\begin{equation}
\iint dxdy \Phi^+(x,y)\Phi(x,y)=\frac12.
\label{5aa}
\end{equation}

According to the asymptotic methods in \cite{VKB},
in the limit as $N\to\infty$,
an asymptotic series of eigenfunctions and eigenvalues 
of the operator~(\ref{3aa}) 
corresponds to each solution of the equations
\begin{equation}
\Omega\Phi(x,y)=\frac{\delta\cH}{\delta\Phi^+(x,y)},
\qquad
\Omega\Phi^+(x,y)=\frac{\delta\cH}{\delta\Phi(x,y)}
\label{6aa}
\end{equation}
that also satisfy condition~(\ref{5aa}). 
It follows from the explicit form of the true symbol~(\ref{4aa})
that system~(\ref{6aa}) can be written as 
\begin{eqnarray}
&&\Omega\Phi(x,y)=-\frac{\hbar^2}{2m}(\Delta_x+\Delta_y)\Phi(x,y)+\nn\\
&&+\iint dx'dy'\left(NV\left(N^{1/3}(x-y)\right)+NV\left(N^{1/3}(x'-y')\right)
\right)\Phi^+(x',y')\Phi(x,x')\Phi(y,y'),\nn\\
&&\Omega\Phi^+(x,y)=-\frac{\hbar^2}{2m}(\Delta_x+\Delta_y)\Phi^+(x,y)+\nn\\
&&+\iint dx'dy'\left(NV\left(N^{1/3}(x-x')\right)+NV\left(N^{1/3}(y-y')\right)
\right)\Phi(x',y')\Phi^+(x,x')\Phi^+(y,y').
\label{7aa}
\end{eqnarray}

In the limit as $N\to\infty$, system~(\ref{7aa})
supplemented with the condition~(\ref{5aa})
has the family of solutions 
\begin{eqnarray}
&&\Phi^+_{k}(x,y)=\frac1{L_1L_2^2}\cos\left(k(x-y)\right),\nn\\
&&\Phi_{k}(x,y)=\frac1{L_1L_2^2}\sum_{l}\vp_{k,l}\exp\left(il(x-y)\right),
\label{8aa}
\end{eqnarray}
where $k$ and $l$ are three-dimensional vectors of the form 
$$
2\pi\left(\frac{n_1}{L_1},\frac{n_2}{L_2},\frac{n_3}{L_2}\right),
$$
the numbers $n_1$, $n_2$, and $n_3$ are integers, 
the functions $\vp_{k,l}$ in formula~(\ref{2aa}) 
have the form 
\begin{equation}
\vp_{k,l}=\frac1{2V_0}\left(\frac{\hbar^2}{2m}(k^2-l^2)+V_0\pm
\sqrt{\left(\frac{\hbar^2}{2m}(k^2-l^2)+V_0\right)^2-V_0^2}\right),
\label{9aa}
\end{equation}
with the plus sign for $l^2>k^2$ 
and the minus sign for $l^2<k^2$, 
and $V_0$ here and hereafter denotes the expression 
\begin{equation}
V_0=\frac1{L_1L_2^2}\int dx V(x),
\label{10aa}
\end{equation}
in which the integral is taken over the space $\bR^3$. 
The vector $k$ in~(\ref{8aa}) plays the role of a parameter
indexing different solutions of system~(\ref{7aa}) with
conditions~(\ref{5aa}). 
The solutions~(\ref{8aa}) are standing waves 
associated with the series without fluidity.

The leading asymptotic term for the eigenvalues in the series 
corresponding to solution~(\ref{8aa}) is equal to the value of
the product of the symbol~(\ref{4aa}) on the functions~(\ref{8aa})
times~$N$, 
\begin{equation}
E_k=N\left(\frac{\hbar^2k^2}{2m}+\frac{V_0}2\right).
\label{11aa}
\end{equation}

The asymptotic expressions for the eigenvalues and
eigenfunctions, in particular, the asymptotic terms
following~(\ref{11aa}) 
are determined by the solutions of system~(\ref{7aa})
and also by the solutions of the variational system of
equations corresponding to~(\ref{7aa}).
The detailed study of solutions of the variational system 
is here omitted. In what follows, we present the results of this
study obtained, in particular, by using the Maple program.

If $k=0$, then the asymptotic series related 
to this solution~(\ref{8aa}) of system~(\ref{7aa}), (\ref{5aa}) 
is the Bogoliubov series corresponding to the ground state
without fluidity. 
The quasiparticle spectrum for this series is expressed by the
well-known formula 
\begin{equation}
\lambda_l=\sqrt{\left(\frac{\hbar^2l^2}{2m}+V_0\right)^2-V_0^2}.
\label{12aa}
\end{equation}

If $k\ne0$, then the variational system equations corresponding
to~(\ref{7aa}) gives the quasiparticle spectrum
\begin{eqnarray}
&&\lambda_{1,k,l}=  \nn \\
&&=\pm\frac{\hbar^2}{2m}\sqrt{k^4+\frac{l^2}2+\frac{l_1^2}2-
k^2l^2-k^2l_1^2+\frac12(l^2+l_1^2-2k^2)\sqrt{(l^2-l_1^2)^2+
\left(\frac{4mV_0}{\hbar^2}\right)^2}},\nn\\
&&\lambda_{2,k,l}= \nn \\
&&=\pm\frac{\hbar^2}{2m}\sqrt{k^4+\frac{l^2}2+\frac{l_1^2}2-
k^2l^2-k^2l_1^2-\frac12(l^2+l_1^2-2k^2)\sqrt{(l^2-l_1^2)^2+
\left(\frac{4mV_0}{\hbar^2}\right)^2}}, \label{13aa}
\end{eqnarray}
where $l_1=l+2k$ and $l\ne k$. 
In the formula for $\lambda_{1,k,l}$ 
the plus sign is chosen for $l^2>k^2$ 
and the minus sign for $l^2<k^2$;
in the formula for $\lambda_{2,k,l}$,
the plus sign is chosen for $l_1^2>k^2$ 
and the minus sign for $l_1^2<k^2$.

The explicit form~(\ref{13aa}) implies that 
this quasiparticle spectrum contains some negative values,
which means that the series corresponding to the solution~(\ref{8aa}) 
with $k\ne0$ is nonmetastable.
In this connection, we suggest the following explanation for the
dependence of the critical velocity on capillary width. 
In what follows, we assume that $L_1\gg L_2$. 
We consider the Bogoliubov series corresponding to the flow 
with the velocity $\hbar k_0/m$ along the capillary, 
where $k_0=2\pi(n_1/L_1,0,0)$. 
For the boson system in question, the leading asymptotic term 
for the eigenvalues in this series is 
\begin{equation}
N\left(\frac{\hbar^2k_0^2}{2m}+\frac{V_0}2\right).
\label{14aa}
\end{equation}
We now assume that there is a relation between $L_1$ and $L_2$
such that there exists a vector $k=2\pi(0,n_2/L_2,n_3/L_2)$
for which the corresponding value of the term in~(\ref{11aa})
is exactly equal to~(\ref{14aa}). This means that there can be 
resonance between the current states of the Bogoliubov series 
and the states of the nonmetastable series 
corresponding to~(\ref{8aa}). 
If the value of $L_1$ is very large, then resonance is also
possible for the case where the value of the term
in~(\ref{11aa})  is close to expression~(\ref{14aa})
but need not coincide with it.
The existence of such a resonance indicates the possibility 
of a transition from the current state to a nonmetastable state
from which the system falls to the lowest energy level, 
which indicates the loss of superfluidity.

The following versions of series of two-particle solutions
are determined by the relation:
\begin{equation}
Q(k_1,k_2,m_1,m_2,m_3,l_1,l_2,l_3)=\frac{l_3^2-m_3^2}
{m_1^2+m_2^2+k_1^2+k_2^2-l_1^2-l_2^2}, \label{1num}
\end{equation}
where $k_1,k_2,l_1,l_2,l_3,m_1,m_2,m_3$ are integers.
We impose the following conditions on the numbers
$k_1,k_2,l_1,l_2,l_3,m_1,m_2,m_3$:
\begin{equation}
Q>0, \qquad k_1^2+k_2^2>0,\label{2num}
\end{equation}
where $Q$ is an irreducible fraction, 
i.e., its numerator and denominator do not have  
common divisors. 

Professor A.~A.~Karatsuba kindly calculated for me 
the frequency of the spectrum of two-particle solutions 
in the interval $[a, a+h]$.
The asymptotic expression in the limit as $L\to\infty$
has the following form.

We are given the functions~(\ref{1num}) and the function
\begin{equation}
V=\sqrt{l^2_1+l^2_2+\frac{l^2_3}{Q}}, \label{3num}
\end{equation}
where $k_1,k_2,m_1,m_2,m_3,l_1,l_2,l_3$ are nonnegative integers
that do not exceed $L$; moreover, $Q>0$ and $k^2_1+k^2_2>0$.

We are also given the numbers $a\geq1$, $0<h\leq 1$.

Let $K$ be the number of sets $k_1,k_2,m_1,m_2m_3,l_1,l_2,l_3$
satisfying the inequality 
\begin{equation}%1
a<V\leq a+h.
\end{equation}
Then $K$ is determined by the approximate formula 
\begin{equation}
K=\bigg(\frac15+\frac19\bigg)\frac{\pi^3}{128}L^2((a+h)^6-a^6)
=\frac{7\pi^3}{2880}L^2((a+h)^6-a^6). \label{4num}
\end{equation}
If the distance between the spectrum points is of the order of 
$\frac 1N$, i.e., if it is determined with the desired accuracy,
then the Bogoliubov series is in resonance.

This readily shows that $a \sim 4$, 
and hence resonance occurs approximately at the points 
$8\pi h/mL_2$. 
This resonance is braking the fluid, 
and the superfluidity disappears.

We here present A.~A.~Karatsuba's proof of formula~(\ref{4num}).

We introduce the new notation 
$$
r=m^2_1+m^2_2+k^2_1+k^2_2,\qquad l=l^2_1+l^2_2,\qquad
\xi=\frac{m_3}{l_3}\geq0.
$$
In this notation, we have 
$$
Q=l^2_3\frac{1-\xi^2}{r-l}>0.
$$
Two cases, (I) and (II), are possible: 
$$
\text{(I)}\qquad 0\leq \xi<1, \ l<r; \qquad \text{(II)}\qquad
\xi>1,\  l>r.
$$
The function $V$ can be rewritten as 
$$
V=\sqrt{l+\frac{r-l}{1-\xi^2}} =\sqrt{\frac{r-l\xi^2}{1-\xi^2}}
$$
and hence~(\ref{4num}) becomes 
\begin{equation}%2
a^2<V^2=\frac{r-l\xi^2}{1-\xi^2}\leq (a+h)^2. \label{5num}
\end{equation}

{\bf{Case (I)}.} From~(\ref{5num}) we obtain 
\begin{eqnarray}%3
&& a^2(1-\xi^2)+l\xi^2<r\leq (a+h)^2(1-\xi^2)+l\xi^2, \nn \\
&&l<r,\qquad 0\leq \xi<1. \label{6num}
\end{eqnarray}

If in~(\ref{6num}) we have the relation 
$$
a^2(1-\xi^2)+l\xi^2\leq l,\qquad\text{i.e.},\quad a^2\leq l,
$$
then system~(\ref{6num}) is equivalent to 
\begin{eqnarray}%4
&& l<r\leq (a+h)^2(1-\xi^2)+l\xi^2, \nn \\
&&a^2\leq l<(a+h)^2, \nn \\
&&0\leq \xi<1. \label{7num}
\end{eqnarray}
But if in~(\ref{6num}) we have 
$$
a^2(1-\xi^2)+l\xi^2> l,\qquad\text{i.e.},\quad l< a^2,
$$
then system~(\ref{6num}) is equivalent to 
\begin{eqnarray}%5
&& a^2(1-\xi^2)+l\xi^2<r\leq (a+h)(1-\xi^2)+l\xi^2, \nn \\
&& l< a^2, \nn \\
&& 0\leq \xi<1.  \label{8num}
\end{eqnarray}
Let $K_1$ be the number of sets satisfying 
either~(\ref{7num}) or~(\ref{8num}).

It follows from systems~(\ref{7num}) and~(\ref{8num}) that 
$$
l<a^2 \quad\text{or}\quad l<(a+h)^2,
$$
i.e., we always have $l_1,l_2<a+h\leq a+1$, since $h\leq 1$.

Precisely in the same way we obtain
$$
r<(a+h)^2,\qquad\text{i.e.}, \quad m_1,m_2,k_1,k_2<a+h\leq a+1.
$$
Therefore, the six variables $l_1,l_2,m_1,m_2mk_1,k_2$ 
are bounded by the value $a+1$, 
and they play an unimportant role for small~$a$.
The main parameter is $L$, $L\to+\infty$, 
and this parameter is related only to the variable~$\xi$,
$$
\xi=\frac{m_3}{l_3}<1,\qquad 0\leq m_3<l_3\leq L.
$$

The ranges of these variables are determined automatically 
by the first two inequalities in systems~(\ref{7num})
and~(\ref{8num}). 
In what follows, we present two asymptotic formulas 
(in the limit as $x\to\infty$).

The number of sets $m_1,m_2,k_1,k_2$
for which $r=m^2_1+m^2_2+k^2_1+k^2_2\leq x$ 
is equal to 
$$
\frac1{24}v_4(\sqrt{x}),
$$
where $v_4(\sqrt{x})$ is the volume of a four-dimensional sphere
of radius $\sqrt{x}$, i.e., 
\begin{equation}%6
v_4(\sqrt{x}) =(\sqrt{x})^4\frac{\pi^{4/2}}{\Gamma(4/2+1)}
=x^2\frac{\pi^2}{2}; \label{9num}
\end{equation}
the number of sets $l_1,l_2$ for which $ l^2_1+l^2_2\leq x$
is equal to 
\begin{equation}%7
\frac14\pi x. \label{10num}
\end{equation}
These are approximate formulas; the larger $x$, 
the more precise they are.

From~(\ref{9num}) and~(\ref{7num}),~(\ref{8num}) we obtain 
\begin{align*}
K_1&=\sum_{0\leq\xi<1} \sideset{}{'}\sum_{a^2\leq l<(a+h)^2}
\frac{\pi^2}{32} \{ ( (a+h)^2(1-\xi^2)+l\xi^2 )^2-(l)^2 \}
\\
&\quad +\sum_{0\leq\xi<1} \sideset{}{'}\sum_{ l<a^2 }
\frac{\pi^2}{32} \{
((a+h)^2(1-\xi^2)+l\xi^2)^2-(a^2(1-\xi^2)+l\xi^2)^2) \}.
\end{align*}
The prime on the sum over $l$ means that 
the numbers~$l$ are regarded with multiplicity taken into
account  
(the number of solutions of the equation is $l=l^2_1+l^2_2$,
$l_1\geq0$, $l_2\geq0$); 
the multiplicity is equal to $\pi/4$ 
(averaged over~$l$), i.e., 
\begin{align*}
K_1&=\sum_{0\leq\xi<1}\sum_{a^2\leq l<(a+h)^2} \frac{\pi^2}{128}
\{ ((a+h)^2(1-\xi^2)+l\xi^2)^2-l^2\}
\\
&\quad +\sum_{0\leq\xi<1} \sum_{ l<a^2 } \frac{\pi^2}{128} \{
((a+h)^2(1-\xi^2)+l\xi^2)^2 -(a^2(1-\xi^2)+l\xi^2)^2) \}
\end{align*}

The sum over~$l$ (with a good accuracy) is equal to the integral
\begin{align*}
K_1&=\frac{\pi^3}{128}\sum_{0\leq\xi<1} \int^{(a+h)^2}_{a^2} 
\{((a+h)^2(1-\xi^2)+l\xi^2)^2-l^2 \}\,dl
\\
&\quad +\frac{\pi^3}{128}\sum_{0\leq\xi<1} \int^{a^2}_{0} 
\{((a+h)^2(1-\xi^2)+l\xi^2)^2-(a^2(1-\xi^2)+l\xi^2)^2 \}\,dl
\\
&=\frac{\pi^3}{128}((a+h)^6-a^6)
\sum_{0\leq\xi<1}\bigg(\frac23-\xi^2+\frac13\xi^4\bigg).
\end{align*}
Since 
$$
\xi=\frac{m_3}{l_3},\qquad 0\leq m_3<l_3\leq L,
$$
the sum over $\xi$ (with a good accuracy) is equal to the
integral 
$$
\int^L_0\int^l_0 \bigg(\frac23-\bigg(\frac{m}{l}\bigg)^2
+\frac13\bigg(\frac{m}{l}\bigg)^4\bigg)\,dm dl
=\frac{L^2}{2}\bigg(\frac23-\frac13+\frac1{15}\bigg)
=\frac{L^2}{5}.
$$
Hence we obtain 
$$
K_1=\frac{\pi^3}{128}((a+h)^6-a^6)\frac{L^2}{5}.
$$

{\bf{Case (II).}}
$$
\xi>1, \qquad l>r, \qquad\text{i.e.},\quad l_3<m_2,\quad r<l.
$$
Let
$$
0<\xi_1=\frac1\xi=\frac{l_3}{m_3}<1, \qquad r<l.
$$
Then we have the equality 
$$
V=\sqrt{\frac{r-l\xi^2}{1-\xi^2}}
=\sqrt{\frac{l-r\xi^2_1}{1-\xi^2_1}}
$$
and the condition $a<V\leq(a+h)$ implies
$$
a^2<\frac{l-r\xi^2_1}{1-\xi^2_1}\leq (a+h)^2
$$
or 
\begin{eqnarray}%8
&&a^2(1-\xi^2_1)+r\xi^2_1 < l \leq (a+h)^2(1-\xi^2_1)+r\xi^2_1,
\nn \\
&&r<l, \qquad 0<\xi_1<1.\label{11num}
\end{eqnarray}

It follows from~(\ref{11num}) that $r<(a+h)^2$. 
If $r>a^2(1-\xi^2_1)+r\xi^2_1$, i.e., if $r>a^2$,
then~(\ref{11num}) is replaced by the system 
\begin{eqnarray}%9
&& r-l<l\leq (a+h)^2(1-\xi^2_1)+r\xi^2_1, \nn \\
&&a^2<r<(a+h)^2, \qquad 0<\xi_1<1. \label{12num}
\end{eqnarray}
If $r\lq a^2(1-\xi^2_1)+r\xi^2_1$, 
i.e., if $r\leq a^2$, then~(\ref{11num}) is replaced by~(\ref{13num}):
\begin{eqnarray}%10
&&a^2(1-\xi^2_1)+r\xi^2_1<l\leq (a+h)^2(1-\xi^2_1)+r\xi^2_1, \nn \\
&&r\leq a^2,  \qquad 0<\xi_1<1.\label{13num}
\end{eqnarray} 
It follows from Case (I) and systems~(\ref{7num})
and~(\ref{8num}) that $\xi$ is replaced by $\xi_1$, 
while $r$ and $l$ interchange.

Therefore, we use formulas~(\ref{10num}),~(\ref{12num}), 
and~(\ref{13num}) to find~$K_2$, which is the number of sets 
satisfying~(\ref{12num}) and~(\ref{13num}):
\begin{align*}
K_2&=\frac14\pi\sum_{0<\xi_1<1} \sideset{}{'}\sum_{a^2<r<(a+h)^2}
((a+h)^2(1-\xi^2_1)+r\xi^2_1-r)
\\
&\quad +\frac14\pi\sum_{0<\xi_1<1} \sideset{}{'}\sum_{r\leq a^2}
((a+h)^2-a^2)(1-\xi^2_1).
\end{align*}
The prime on the sum over $r$ means that the number of
solutions~$r$ is regarded with multiplicity taken into account 
(the number of solutions of the equation is 
$r=m^2_1+m^2_2+k^2_1+k^2_2$).

The number of solutions of the inequality 
$$
m^2_1+m^2_2+k^2_1+k^2_2\leq x,\qquad m_1,m_2,k_1,k_2\geq 0,.
$$
is approximately equal to 
$$
\frac{\pi^2}{32}x^2.
$$
We obtain 
\begin{align*}
&\sideset{}{'}\sum_{a^2<r<(a+h)^2} r
=\sideset{}{'}\sum_{a^2<r<(a+h)^2} \sum_{0\leq t<r} 1 =\sum_{0\leq
t<(a+h)^2} \sideset{}{'}\sum_{\max(a^2,t)<r<(a+h)^2} 1
\\
&\qquad =\sum_{0\leq t\leq a^2} \sideset{}{'}\sum_{a^2<r<(a+h)^2}
1 + \sum_{a^2< t< (a+h)^2} \sideset{}{'}\sum_{t<r<(a+h)^2} 1
\\
&\qquad =\frac{\pi^2}{32} \bigg(\sum_{0\leq t\leq (a+h)^2}
((a+h)^4-a^4) + \sum_{a^2< t< (a+h)^2} ((a+h)^4-t^2)\bigg)
\\
&\qquad =\frac{\pi^2}{32}
\bigg((a+h)^4a^2-a^6+(a+h)^4((a+h)^2-a^2)
-\frac13((a+h)^6-a^6)\bigg)
\\
&\qquad =\frac{\pi^2}{3\cdot 16}((a+h)^6-a^6);
\end{align*}
We derive the following formula for $K_2$:
\begin{align*}
K_2&=\frac{\pi}{4}\sum_{0<\xi_1<1}\bigg\{
(a+h)^2(1-\xi^2_1)\frac{\pi^2}{32}(()^4-a^4)
\\
&\qquad -(1-\xi^2_1)\frac{\pi^2}{3\cdot 16}((a+h)^6-a^6)
+((a+h)^2-a^2)(1-\xi^2_1)\frac{\pi^2}{32}a^4\bigg\}
\\
&=\frac{\pi^3}{12\cdot 32}((a+h)^6-a^6)
\sum_{0<\xi_1<1}(1-\xi^2_1).
\end{align*}
Since $\xi_1=\frac{l}{m}$, $0<l<m\leq L$, we have 
$$
\sum_{0<\xi_1<1}(1-\xi^2_1)
=\int^L_0\int^m_0\bigg(1-\bigg(\frac{l}{m}\bigg)^2\bigg)\,dl dm
=\frac{L^2}{3}
$$
and
$$
K_2=\frac{\pi^3}{32\cdot 36}L^2( (a+h)^6-a^6 ).
$$
The fact that $K=K_1+K_2$ implies the relation
$$
K=\bigg(\frac15+\frac19\bigg)\frac1{128}\pi^3 L^2( (a+h)^6-a^6 )
=\frac{7\pi^3}{2880}L^2( (a+h)^6-a^6 ).
$$

{\bf Remarks.}

1. The formula for $K$ is an approximate formula.

2. This formula is sufficiently exact 
if $L$ is large and $a$ is not too small but significantly less than~$L$; 
for example, $a=\sqrt{L}$, \  $L\to+\infty$.

3. If $a$ is chosen to be not large, for example, 
$a=4$, $h=\frac12$, or $a=3$, $h=\frac12$, 
or $a=5$, $h=\frac12$,  
then we can obtain sufficiently exact formulas for $L\geq8$, 
since, in this case, 
we can exactly calculate the multiplicities of $l$ and 
the $l$ themselves (as sums of two squared numbers).
But, in this case, we must perform specific calculations.

{\bf Example of calculations for small $a$.}

Let $a=4$, $h=1$; $L>5$.

In system~(\ref{7num}), $l$ satisfy the condition
$$
16\leq l=l^2_1+l^2_2<25.
$$
The pairs $(l_1,l_2)$ satisfying this condition are:
$$
(0,4), (1,4), (2,4), (3,3), (4,1), (4,1), (4,0),
$$ 
i.e., from~(\ref{7num}) we obtain two systems for $l=16$,
$l=17$, and $l=20$ and one system for $l=18$:
\begin{eqnarray}%11
&& 16<r\leq 25-9\xi^2, \nn \\
&& 0\leq\xi<1, \label{14num}
\end{eqnarray}
\begin{eqnarray}%12
&& 17<r\leq 25-8\xi^2, \nn \\
&& 0\leq\xi<1,\label{15num}
\end{eqnarray}
\begin{eqnarray}%13
&& 20<r\leq 25-5\xi^2, \nn \\
&& 0\leq\xi<1,\label{16num}
\end{eqnarray}
\begin{eqnarray}%14
&& 18<r\leq 25-7\xi^2, \nn \\
&& 0\leq\xi<1.\label{17num}
\end{eqnarray}

To calculate the number of sets $(m_1,m_2,k_1,k_2)$
for which $r=m^2_1+m^2_2+k^2_1$ satisfies 
systems~(\ref{14num})--(\ref{17num}),
we can either use the computer software
(in this case, 
we must divide $\xi$ into intervals, 
where $9\xi^2$ lies between two neighboring integers 
from~$0$ to~$9$, the same for $8\xi^2$, $5\xi^2$, and $7\xi^2$,
and calculate the number of sets directly)
or use the formula for the number of integer points 
in the four-dimensional ball 
and thus find the points with positive coordinates. 
This is an approximate formula and if 
$$
m^2_1+m^2_2+k^2_1\leq x,\qquad m_i,k_i\geq0,
$$
then this number is equal to 
\begin{equation}%15
\frac{\pi^2}{32}x^2. \label{18num}
\end{equation}
Therefore, for system~(\ref{7num}), we obtain the number of sets
\begin{align}%16
&\frac{\pi^2}{32}\sum_{0\leq\xi<1} \Big\{((25-9\xi^2)^2-16^2)\cdot
2 +((25-8\xi^2)^2-17^2)\cdot 2
\nonumber\\
&\qquad\qquad +((25-5\xi^2)^2-20^2)\cdot 2
+((25-7\xi^2)^2-18^2)\Big\}. \label{19num}
\end{align}
Since the number of such $l$ is equal to~$2$, 
we use the factor~$2$. In system~(\ref{8num}),
we have 
$$
l=l^2_1+l^2_2<16,
$$
and the pairs $(l_1,l_2)$ are: 
$\underline{(0,3)}$, $\underline{(0,2)}$,
$\underline{(0,1)}$, $\underline{(0,0)}$, 
$\underline{(1,2)}$, $\underline{(1,3)}$, 
$\underline{(1,0)}$, $\underline{(2,0)}$,
$\underline{(2,1)}$, $(2,2)$, $(2,3)$, 
$\underline{(3,0)}$, $\underline{(3,2)}$, 
$(3,2)$.

\begin{center}
\begin{tabular}{|c|c|c|c|c|c|c|c|c|}\hline
l=        & 0 & 1 & 4 & 9 & 5 & 10 & 8 & 13 \\ \hline 
multiplicity & 1 & 2 & 2 & 2 & 2 & 2  & 1 & 2  \\ \hline
\end{tabular}
\end{center}

System~(\ref{8num}) for each~$l$ has the form 
$$
16(1-\xi^2)+l\xi^2 < r \leq 25(1-\xi^2)+l\xi^2,\\
0\leq\xi <1.
$$
Using~(\ref{18num}), we obtain the following formula 
(with multiplicity of $l$ taken into account)
for the number of sets for system~(\ref{8num}):
\begin{align}%17
&\frac{\pi^2}{32}\sum_{0\leq\xi<1}\Big\{
(25(1-\xi^2))^2-(16(1-\xi^2))^2
\nonumber\\
&\qquad\qquad +( (25(1-\xi^2)+\xi^2)^2-(16(1-\xi^2)+\xi^2)^2
)\cdot 2
\nonumber\\
&\qquad\qquad +( (25(1-\xi^2)+4\xi^2)^2-(16(1-\xi^2)+4\xi^2)^2
)\cdot 2
\nonumber\\
&\qquad\qquad +( (25(1-\xi^2)+9\xi^2)^2-(16(1-\xi^2)+9\xi^2)^2
)\cdot 2
\nonumber\\
&\qquad\qquad +( (25(1-\xi^2)+5\xi^2)^2-(16(1-\xi^2)+5\xi^2)^2
)\cdot 2
\nonumber\\
&\qquad\qquad +( (25(1-\xi^2)+10\xi^2)^2-(16(1-\xi^2)+10\xi^2)^2
)\cdot 2
\nonumber\\
&\qquad\qquad +( (25(1-\xi^2)+8\xi^2)^2-(16(1-\xi^2)+8\xi^2)^2
)\cdot 1
\nonumber\\
&\qquad\qquad +( (25(1-\xi^2)+13\xi^2)^2-(16(1-\xi^2)+13\xi^2)^2
)\cdot 2 \Big\}
\nonumber\\
& =\frac{\pi^2}{32}\sum_{0\leq\xi<1}\Big\{
(25(1-\xi^2))^2-(16(1-\xi^2))^2
\nonumber\\
&\qquad\qquad +((25-24\xi^2)^2-(16-15\xi^2)^2)\cdot 2
\nonumber\\
&\qquad\qquad +((25-21\xi^2)^2-(16-12\xi^2)^2)\cdot 2
\nonumber\\
&\qquad\qquad +((25-16\xi^2)^2-(16-7\xi^2)^2)\cdot 2
\nonumber\\
&\qquad\qquad +((25-20\xi^2)^2-(16-11\xi^2)^2)\cdot 2
\nonumber\\
&\qquad\qquad +((25-15\xi^2)^2-(16-6\xi^2)^2)\cdot 2
\nonumber\\
&\qquad\qquad +((25-17\xi^2)^2-(16-8\xi^2)^2)\cdot 1
\nonumber\\
&\qquad\qquad +((25-12\xi^2)^2-(16-3\xi^2)^2)\cdot 2 \Big\}.
\label{20num}
\end{align}
Since  
$\xi=\frac{m_3}{l_3}$, $0\leq m_3\leq l_3$, $1\leq l_3\leq L$, 
the number of sets corresponding to Case~(I)
is either calculated directly according
to~(\ref{19num})--(\ref{20num}) 
or is approximately replaced by an integral of the form 
$$
\int^L_0 dl \int^l_0 dm \dots
$$
equal to 
$$
\frac{\pi^2}{64} L^2 \mu,
$$
where
\begin{align*}
\mu&=\int^1_0 \Big\{ 2( (25-9u^2)^2-16^2 ) + 2( (25-8u^2)^2-17^2 )
\\
&\qquad +2( (25-5u^2)^2-20^2 ) + ( (25-7u^2)^2-18^2)
\\
&\qquad +( 25(1-u^2) )^2-( 16(1-u^2) )^2 +2(
(25-24u^2)^2-(16-15u^2)^2 )
\\
&\qquad +2( (25-21u^2)^2-(16-12u^2)^2 ) +2(
(25-16u^2)^2-(16-7u^2)^2 )
\\
&\qquad +2( (25-20u^2)^2-(16-11u^2)^2 ) +2(
(25-15u^2)^2-(16-6u^2)^2 )
\\
&\qquad +1( (25-17u^2)^2-(16-8u^2)^2 ) +2(
(25-12u^2)^2-(16-3u^2)^2 )\Big\}\,du.
\end{align*}
Thus, we have 
$$
K_1=\frac{\pi^2}{64}L^2\mu.
$$

Case~(II), 
i.e., the calculation of $K_2$ is the same;
only the multiplicity of $r$, 
$16<r<25$, $r=m^2_1+m^2_2+k^2_1+k^2_2$, 
is calculated either by an exhaustive search 
of the sets $(m_1,m_2,k_1,k_2)$ (also for $r<16$) 
or by the formula for the number of integer points 
in the four-dimensional ball;
the number of sets for $l\leq r$ from~(\ref{12num})
and~(\ref{13num}) is calculated approximately 
by the formula 
$$
\frac{\pi}{4}x.
$$
Therefore, if $\kappa(r)$ is the multiplicity of $r$, 
then 
\begin{align*}
K_2&=\sum_{0<\xi_1<1}\bigg\{
\sum_{16<r<25}\kappa(r)\frac{\pi}{4}(25-r)(1-\xi^2_1)
+\sum_{r<16}\frac{\pi}{4}\kappa(r)(25-16)(1-\xi^2_1)\bigg\}= \nn
\\
&=\frac{L^2}{2}\frac23\frac{\pi}{4}
\bigg(\sum_{16<r<25}\kappa(r)(25-r) +9\sum_{r<16}\kappa(r)\bigg).
\nn
\end{align*}

But we have already calculated that 
\begin{align*}
\sum_{r<16}\kappa(r)&=\frac{\pi^2}{32}4^4; \nn \\
\sum_{16<r<25}\kappa(r)&=\frac{\pi^2}{32}(5^4-4^4); \nn \\
\sum_{16<r<25}\kappa(r)r&=\frac{\pi^2}{48}(5^6-4^6). \nn
\end{align*}
Hence we have 
\begin{align*}
K_2&=\frac{\pi}{12}L^2\bigg(25\cdot\frac{\pi^2}{32}(5^4-4^4)
-\frac{\pi^2}{48}(5^6-4^6) +9\frac{\pi^2}{32}\cdot4^4\bigg),\nn
\\
K&=K_1+K_2=cL^2.
\end{align*}
All these formulas are approximate.

The number $\mu=\int^1_0\dots dm$ 
and the constant~$c$ in the formula for~$K$
have not been calculated, but these calculations 
are very simple.

The author wishes to express his deep gratitude to 
Professor A.~A.~Karatsuba and to G.~V.~Koval' for 
their help in preparing this paper.

\end{document}